\documentclass[aps,twocolumn]{revtex4-1}

\usepackage[utf8]{inputenc}
\usepackage[T1]{fontenc}
\usepackage[english]{babel}
\usepackage{graphicx}
\usepackage{latexsym,amsmath,amssymb,amsthm}
\usepackage{makeidx}
\usepackage[usenames,dvipsnames]{color}
\usepackage[unicode=true,colorlinks=true,linkcolor=RoyalBlue,citecolor=RoyalBlue]{hyperref}
\usepackage{natbib}

\newcommand{\hetre}{{^3\text{H\lowercase{e}}}}

\newcommand{\mrtwo}{$\text{M(RT)}^2$ }

\newcommand{\psiswf}{\psi_{\text{\tiny SWF}}}
\newcommand{\psifswf}{\psi_{\text{\tiny FSWF}}}

\newcommand{\hloc}{\text{H}^{\text{loc}}}
\newcommand{\SD}{\text{SD}}

\def \NN {\nonumber \\}

\makeindex

\begin{document}
	
\title{On the Sign Problem of the Fermionic Shadow Wave Function}
\author{Francesco \surname{Calcavecchia}}
\affiliation{Institute of Physics, Johannes Gutenberg-University, Staudingerweg 7, D-55128 Mainz, Germany}
\affiliation{Graduate School Materials Science in Mainz, Staudingerweg 9, D-55128 Mainz, Germany}
\author{Francesco \surname{Pederiva}}
\affiliation{Dipartimento di Fisica, University of Trento, via Sommarive 14, I-38050 Povo, Trento, Italy}
\affiliation{INFN-TIFPA, Trento Institute for Fundamental Physics and Applications, Trento, Italy}
\author{Malvin H. \surname{Kalos}}
\affiliation{Lawrence Livermore National Laboratory Livermore, California 94550, USA}
\author{Thomas D. \surname{K\"uhne}}
\email{tdkuehne@mail.uni-paderborn.de}
\affiliation{Institute of Physical Chemistry and Center for Computational Sciences, Johannes Gutenberg University Mainz, Staudinger Weg 7, D-55128 Mainz, Germany}
\affiliation{Department of Chemistry, University of Paderborn, Warburger Str. 100, D-33098 Paderborn, Germany}
\date{\today}

\begin{abstract}
We present a whole series of novel methods to alleviate the sign problem of the Fermionic Shadow Wave Function in the context of Variational Monte Carlo. 
The effectiveness of our new techniques is demonstrated on the example of liquid $\hetre$. We found that although the variance is substantially reduced, the gain in efficiency is restricted by the increased computational cost. Yet, this development not only extends the scope of the Fermionic Shadow Wave Function, but also facilitates highly accurate Quantum Monte Carlo simulations previously thought not feasible.  
\end{abstract}

\maketitle

\section*{Introduction}
\label{sec:introduction}

The difficulty to solve the Schr\"odinger equation for many interacting particles is because of the fact that it is in general impossible to analytically solve it for more than a few particles. 
Quantum Monte Carlo techniques \cite{QMCentropy2014, QMCrpp2011, QMCjpcm2010}, such as Variational Monte Carlo (VMC) \cite{McMillan:1965uq}, are stochastic methods that allow to numerically solve the many-body Schr\"odinger equation. The main concepts underlying VMC are the application of the Rayleigh-Ritz variational principle and the use of importance sampled Monte Carlo (MC) to efficiently evaluate the high-dimensional integrals of many different expectation values such as the energy \cite{Kalos:MC, Binder:MC}. 
Its great appeal is based upon the low computational complexity, as opposed to wave function based quantum-chemical methods \cite{RevModPhys.71.1267}. Since many-body correlation effects are taken into account by a prescribed trial wave function, VMC is substantially more accurate than commonly employed mean-field techniques, such as Hartree-Fock and density functional theory \cite{HelgakerBook}, and permits to treat even strongly correlated systems. However, since the exact wave function is unknown from the outset, the trial wave function ought to resemble it as closely as possible. Nevertheless, given that the addition of a simple correlation function of the Jastrow form enables to recover most of the correlation effects \cite{PhysRev.98.1479}, VMC typically yields excellent results. 

Here, we consider the Shadow Wave Function (SWF), first introduced by Kalos and coworkers \cite{PhysRevLett.60.1970, PhysRevB.38.4516}, as our trial wave function. The SWF allows to describe all possible condensed phases (gas, liquid and solid) and even phase coexistence within the same functional form \cite{PhysRevLett.72.2589}. Therefore, it is for instance possible to simulate a solid without \textit{a priori} knowing its crystal structure, which instead emerges from the calculation. Moreover, it is even feasible to describe inhomogeneous systems \cite{PhysRevB.56.5909, PhysRevB.69.024203, dandrea:fswf}. In addition, the SWF has further advantageous properties, as for instance that it introduces many-body correlations and obeys a strong similitude with the exact ground state wave function.

Since fermions must obey Fermi-Dirac statistics to comply with the Pauli exclusion principle, an antisymmetric version of the SWF is required that changes the sign upon interchanging any two like-spin particles. 
While these extensions of the SWF to fermionic systems indeed constitute a substantial improvement, when compared to other more conventional trial wave functions, they are plagued by the occurrence of a sign problem, which limits its applicability to rather small systems \cite{calcavecchia:junq_paper}. 
Generally, an efficient and accurate method to simulate fermionic systems thus remains an open and upmost challenging problem. 
In this paper, we therefore study the origin and nature of the sign problem and present multiple of novel methods to alleviate it. 

The remaining of the paper is organized as follows. In section \ref{sec:the_shadow_wave_function} we introduces the SWF and its antisymmetric extension, while in \ref{sub:the_sign_problem} the associated sign problem of the latter is described. Sections \ref{sec:antithetic_contributions} and \ref{sec:the_grouping_technique} describes two kinds of novel approaches to reduce the sign problem, whereas in section \ref{sec:discussion} all the methods presented in the previous sections are assessed in terms of their efficiency. The last section contains the conclusions.

\section{The Shadow Wave Function}
\label{sec:the_shadow_wave_function}

Let us begin by defining the SWF that is obtained by introducing auxiliary degrees of freedom $\mathbf{S} = \left( \mathbf{s}_1, \mathbf{s}_2, \dots \mathbf{s}_N \right)$, called shadows, and integrating over all of them \cite{PhysRevLett.60.1970}. 
In its general form the SWF reads as 
\begin{equation}
	\psiswf(\mathbf{R}) = \int d\mathbf{S} \, \Gamma(\mathbf{R}, \mathbf{S}),
\end{equation}
where $\Gamma(\mathbf{R}, \mathbf{S})$ is an arbitrary wave function, while $\mathbf{R}=\left( \mathbf{r}_1, \mathbf{r}_2, \dots \mathbf{r}_N \right)$ represents all $N$ particle coordinates.

However, the extension of the SWF to fermionic systems is non-trivial, due to the antisymmetry requirement of the wave function to obey the Pauli exclusion principle.
The simplest \textit{ansatz} to achieve this, is known as the Antisymmetric Shadow Wave Function (ASWF) \cite{PhysRevB.53.15129}
\begin{equation}
	\psi_{\text{\tiny ASWF}}(\mathbf{R}) \equiv \text{SD}(\mathbf{R}) J_p(\mathbf{R}) \int d\mathbf{S} \, \Xi(\mathbf{R},\mathbf{S}) J_s (\mathbf{S}),
\end{equation}
where $\text{SD}(\mathbf{R})$ is a Slater determinant that satisfies the antisymmetry condition by changing sign upon the exchange of any two fermions \cite{PhysRev.34.1293}. 
Two-body correlations between the particles are taken into account by a Jastrow correlation factor $J_p(\mathbf{R}) = e^{- \frac{1}{2} \sum_{i<j} u_{pp}(|\mathbf{r}_i - \mathbf{r}_j|)}$ \cite{PhysRev.98.1479} and likewise interactions between the shadows are introduced via $J_s(\mathbf{S}) = e^{- \sum_{i<j} u_{ss}(|\mathbf{s}_i - \mathbf{s}_j|)}$. The kernel $\Xi(\mathbf{R},\mathbf{S}) = e^{- \sum_{i=1}^{N_p} u_{ps}(|\mathbf{r}_i - \mathbf{s}_i|)}$ is to connect the particles with the shadows and can also be interpreted as a Green's function. 
Therein, $u_{pp}$, $u_{ss}$ and $u_{ss}$, respectively, are denoted as two-body pseudopotentials because of their similarity to the potential in the Boltzmann distribution.
Here, we have employed 
\begin{subequations}
\begin{eqnarray}
	u_{pp}(\mathbf{r}_{ij}) &=& \left(\frac{b}{|\mathbf{r}_i - \mathbf{r}_j|}\right)^5 \label{upp} \\
	u_{ss}(\mathbf{s}_{ij}) &=& c_1 V(c_2 |\mathbf{s}_i - \mathbf{s}_j|) \label{uss} \\
	u_{ps}(|\mathbf{r}_i - \mathbf{s}_i|) &=& C |\mathbf{r}_i - \mathbf{s}_i|^2 \label{ups},
\end{eqnarray}
\end{subequations}
where $V$ is the potential used in the Hamiltonian, while $b$, $c_1$, $c_2$ and $C$ are variational parameters.
In order to preserve the translational symmetry of the wave function, which is one of the many appealing properties of the SWF, plane wave orbitals are the natural choice to built up $\text{SD}(\mathbf{R})$. As we have considered an unpolarized system, we have adopted a product of two Slater determinants to describe spin-up and spin-down atoms, respectively, i.e. $\text{SD}(\mathbf{R}) = \text{SD}^{\uparrow}(\mathbf{R}^{\uparrow}) \times \text{SD}^{\downarrow}(\mathbf{R}^{\downarrow})$.

The advantage of the ASWF with respect to conventional trial wave functions is that many-body correlation effects of any order are included from the outset. In fact, even if the shadows are correlated through a two-body function only, the convolution integral permits even higher-order correlation effects to be taken into account. In particular at the presence of phase transitions, where these subtle many-body correlation effects are essential, the ASWF has proven to be superior to ordinary trial wave functions \cite{PhysRevB.53.15129}. 
However, only symmetric correlation effects are taken in account, whereas backflow correlation is not considered \cite{PhysRev.94.262, PhysRev.102.1189, PhysRevB.19.2504, PhysRevLett.46.728, PhysRevLett.47.807, PhysRevB.52.13547, PhysRevB.53.15129, PhysRevB.58.6800, PhysRevB.74.104510}. Furthermore, the nodal surface is imposed \textit{a priori} by a single Slater determinant, and as such only improvable within the flexibility of $\text{SD}(\mathbf{R})$.

Nevertheless, a more intriguing way to devise an antisymmetric version of the SWF is to introduce a $\text{SD}$ as a function of $\mathbf{S}$.
The resulting Fermionic Shadow Wave Function (FSWF) \cite{KalosReatto1995, PederivaChester1998, calcavecchia:junq_paper} reads as 
\begin{equation}
	\psi_{\text{\tiny FSWF}}(\mathbf{R}) = J_p(\mathbf{R}) \int d\mathbf{S} \, \Xi(\mathbf{R}, \mathbf{S}) \text{SD}(\mathbf{S}) J_s(\mathbf{S}).  \label{eq:psi_fswf_naive}
\end{equation}
In fact, given an arbitrary like-spin odd-particle permutation operator $\mathcal{P}$, and exploiting that $\Xi(\mathcal{P} \mathbf{R}, \mathbf{S})=\Xi(\mathbf{R},\mathcal{P} \mathbf{S})$,
\begin{eqnarray}
	\psifswf(\mathcal{P}\mathbf{R}) &=& J_p(\mathcal{P} \mathbf{R}) \int d\mathbf{S} \, \Xi(\mathcal{P} \mathbf{R}, \mathbf{S}) \text{SD}(\mathbf{S}) J_s(\mathbf{S}) \nonumber \\
	 &=& J_p(\mathbf{R}) \int d\mathbf{S} \, \Xi(\mathbf{R}, \mathcal{P} \mathbf{S}) \text{SD}(\mathbf{S}) J_s(\mathbf{S}) \nonumber \\
	 &=& J_p(\mathbf{R}) \int d(\mathcal{P} \mathbf{S}) \, \Xi(\mathbf{R}, \mathcal{P} \mathbf{S}) \left(-\text{SD}(\mathcal{P} \mathbf{S}) \right) \nonumber \\
	 && \times \, J_s(\mathcal{P} \mathbf{S}) \nonumber \\
	 & = & - \psifswf(\mathbf{R}).
\end{eqnarray}
The FSWF has several advantages over the ASWF: (i) It closer resembles the projection onto the exact fermionic ground state (no one actually knows how the propagator on the lowest antisymmetric state behaves), (ii) in the limits of high and low density, the exact asymptotic nodal structure is correctly reproduced and (iii) backflow correlation effects are naturally included \cite{KalosReatto1995}. 

\begin{table}
\begin{equation*}
\begin{array}{|c|c|c|}
\hline
	\text{Trial wave function}         &  \text{Energy per particle}                                          &  N   \\
\hline
\hline
	\text{J-SD}                     &  -1.004 \pm 0.006~\text{K}             & 66  \\
\hline
	\text{ASWF}                   &  -1.222 \pm 0.006~\text{K}             & 66  \\
\hline
	\text{FSWF}                   &  -1.8 \pm 0.2~\text{K}             & 66  \\
\hline
\end{array}
\end{equation*}
\caption{\label{table:risultati_fswf} The ground state energy per particle of liquid $\hetre$ as obtained by VMC using different trial wave functions. The Jastrow-Slater Determinant (J-SD) trial wave function is defined as $\psi_{\text{J-SD}} \equiv \text{SD}(\mathbf{R}) J_p(\mathbf{R})$.}
\end{table}
As can be seen in Table~\ref{table:risultati_fswf}, the FSWF provides a much improved variational ground state energy of liquid $\hetre$, even though with an admitted large statistical uncertainty. 
The corresponding computational details are given in \footnote{{We considered an unpolarized 3D system of $\hetre$ at a density equal to $0.016588~\text{\AA}^{-3}$ (liquid phase) using the Aziz potential HFDHE2 \cite{Aziz:potential, Sofianos:aziz_potential} and periodic boundary conditions in order to mimic an essentially infinite system. We remark that whenever a SD of simple plane waves $e^{i \mathbf{k} \cdot \mathbf{r}}$ is used, the occurrence of a drift (i.e. $ \Sigma_{\beta} \mathbf{k}_{\beta} \neq 0$), as well as anisotropy has to be explicitly taken into account. The simplest way to remedy this is to consider only magic numbers for $N$, i.e. numbers that fill the momenta shell. For a 3D polarized system, they are $1$, $7$, $19$, $27$, $33$, etc. and for an unpolarized system $2$, $14$, $38$, $54$, $66$, etc. We have set the variational parameters for the SWF to be $b=2.76~\text{\AA}, c_1=0.11~\text{K}^{-1}, c_2=0.88$ and $C=0.55~\text{\AA}^{-2}$, as suggested in \cite{PhysRevB.53.15129}, whereas for the J-SD trial wave function we have employed $b=2.9~\text{\AA}$.} \label{note:liquidHe3}}
But, as we are going to explain in detail in the next section, the FSWF entails a serious sign problem that makes it computationally rather expensive to obtain reliable results for large systems. This system size limitation not only restricts the applicability, but also the reliability of the FSWF, due to the presence of significant finite-size effects. 
Therefore, it would be highly desirable to solve - or at least alleviate - the sign problem, and to facilitate very accurate simulations using the FSWF, though with many more particles than presently feasible.

\section{The Sign Problem of the FSWF}
\label{sub:the_sign_problem}
We will illustrate the sign problem of the FSWF on the example of the ground state energy $E$, which in VMC is estimated by
\begin{equation}
	E \simeq \frac{1}{M} \sum_{i=1}^{M} \hloc(\mathbf{R}_i), 
\end{equation}
where $\hloc(\mathbf{R}_i)=\frac{\text{H} \psi(\mathbf{R}_i)}{\psi(\mathbf{R}_i)}$ is the local energy and $M$ the number of sampling points. To this end, the positions $\mathbf{R}_i$ of the particles are sampled from the probability density function (pdf) $\psi^2(\mathbf{R})$, where $\psi$ is the preassigned trial wave function. Assuming that $\psi$ is real, the required positiveness of $\psi^2(\mathbf{R})$ is satisfied by definition. 

In conjunction with the previously introduced shadows, the energy reads as
\begin{equation}
	E = \frac{\int d\mathbf{R} \, d\mathbf{S}_1 d\mathbf{S}_2 \, \Gamma(\mathbf{R},\mathbf{S}_1) \text{H} \Gamma(\mathbf{R},\mathbf{S}_2)}{\int d\mathbf{R} \, d\mathbf{S}_1 d\mathbf{S}_2 \, \Gamma(\mathbf{R},\mathbf{S}_1) \Gamma(\mathbf{R},\mathbf{S}_2)}, \label{eq:energy_evaluation}
\end{equation}
where $\mathbf{R}_i$, $\mathbf{S}_{1 i}$ and $\mathbf{S}_{2 i}$ should in principle be sampled from the pdf $\Gamma(\mathbf{R},\mathbf{S}_{1}) \times \Gamma(\mathbf{R},\mathbf{S}_{2})$.
But, due to the fact that the FSWF is evaluated using two different shadows, $\text{SD}(\mathbf{S}_1)$ and $\text{SD}(\mathbf{S}_2)$, the necessary positiveness requirement of the sampled function is no longer fulfilled.  
As a consequence, it is not possible to sample $\mathbf{R}_i$, $\mathbf{S}_{1 i}$, and $\mathbf{S}_{2 i}$ directly from the pdf $\Gamma(\mathbf{R}, \mathbf{S}_{1}) \times \Gamma(\mathbf{R}, \mathbf{S}_{2})$.

In spite of that, it is feasible to sample from the pdf $|\Gamma(\mathbf{R},\mathbf{S}_{1}) \times \Gamma(\mathbf{R},\mathbf{S}_{2})|$ by introducing the weights $w(\mathbf{R},\mathbf{S}_1,\mathbf{S}_2) = \text{sign}(\Gamma(\mathbf{R},\mathbf{S}_1) \times \Gamma(\mathbf{R},\mathbf{S}_2))$ and estimating the energy as
\begin{equation}
	E \simeq \frac{\sum_{i=1}^{M} \frac{w_i}{2} \left( \hloc_{1 i}+\hloc_{2 i} \right) }{\sum_{i=1}^{M} w_i } , \label{eq:energy_estimator_sum}
\end{equation}
where
\begin{equation*}
	\hloc_1 \equiv \frac{\text{H} \Gamma(\mathbf{R},\mathbf{S}_1)}{\Gamma(\mathbf{R},\mathbf{S}_1)} \quad \text{and} \quad \hloc_2 \equiv \frac{\text{H} \Gamma(\mathbf{R},\mathbf{S}_2)}{\Gamma(\mathbf{R},\mathbf{S}_2)}.
\end{equation*}
However, due to the sign $w(\mathbf{R}_i,\mathbf{S}_{1 i},\mathbf{S}_{2 i})$, the sum of Eq.~\ref{eq:energy_estimator_sum} is typically very slowly converging. This is particularly severe for disordered systems, such as liquid $\hetre$.

In order to evaluate the mean value of $E$ and its unbiased error bar $\sigma$ we have employed the so-called blocking technique \cite{Kalos:MC, BlockingTechnique}. To that extent, the data set is divided into $n_{\text{block}}$ disjoint blocks (typical values for $n_{\text{block}}$ are between 4 and 50), each one with its corresponding average value $E^{\text{block}}_{j}$. Hence, the average energy $\langle E \rangle_{\text{block}}$ and the corresponding variance $\sigma_{\text{block}}^2$ can be computed as  
\begin{subequations}
\begin{eqnarray}
	\langle E \rangle_{\text{block}} &=& \frac{1}{n_{\text{block}}} \sum_{j=1}^{n_{\text{block}}} E^{\text{block}}_{j}\text{and} \\
	\sigma_{\text{block}}^2           &=& \frac{1}{n_{\text{block}}-1} \sum_{j=1}^{n_{\text{block}}} \left( E^{\text{block}}_{j} - \langle E \rangle_{\text{block}} \right)^2.
\end{eqnarray}
\end{subequations}
The straightforward evaluation of the standard deviation $\sigma$ would provide an biased estimate that may severely underestimate the true error bar due to the presence of serial correlation between successive data points.
Nevertheless, given that the length of each block $\frac{M}{n_{\text{block}}}$ is large enough, serial correlation between the block averages $E^{\text{block}}_{j}$ becomes arbitrarily small with the result that $\sigma$ can after all be correctly estimated. 
In fact, when plotting $\sigma_{\text{block}}$ as a function of $n_{\text{block}}$ and assuming that $M$ is sufficiently large, a plateau that corresponds to the correct estimation of the unbiased error bar is emerging. We remark that, mathematically speaking, $\langle E \rangle_{\text{block}}$ may vary for different values of $n_{\text{block}}$, but as long as $M$ is large enough, each value $E^{\text{block}}_{j}$ will be very close to $E$, so that eventually $\langle E \rangle_{\text{block}}$ will be independent from the choice of $n_{\text{block}}$.

\begin{figure}
	\centering
		\includegraphics[width=8.5cm]{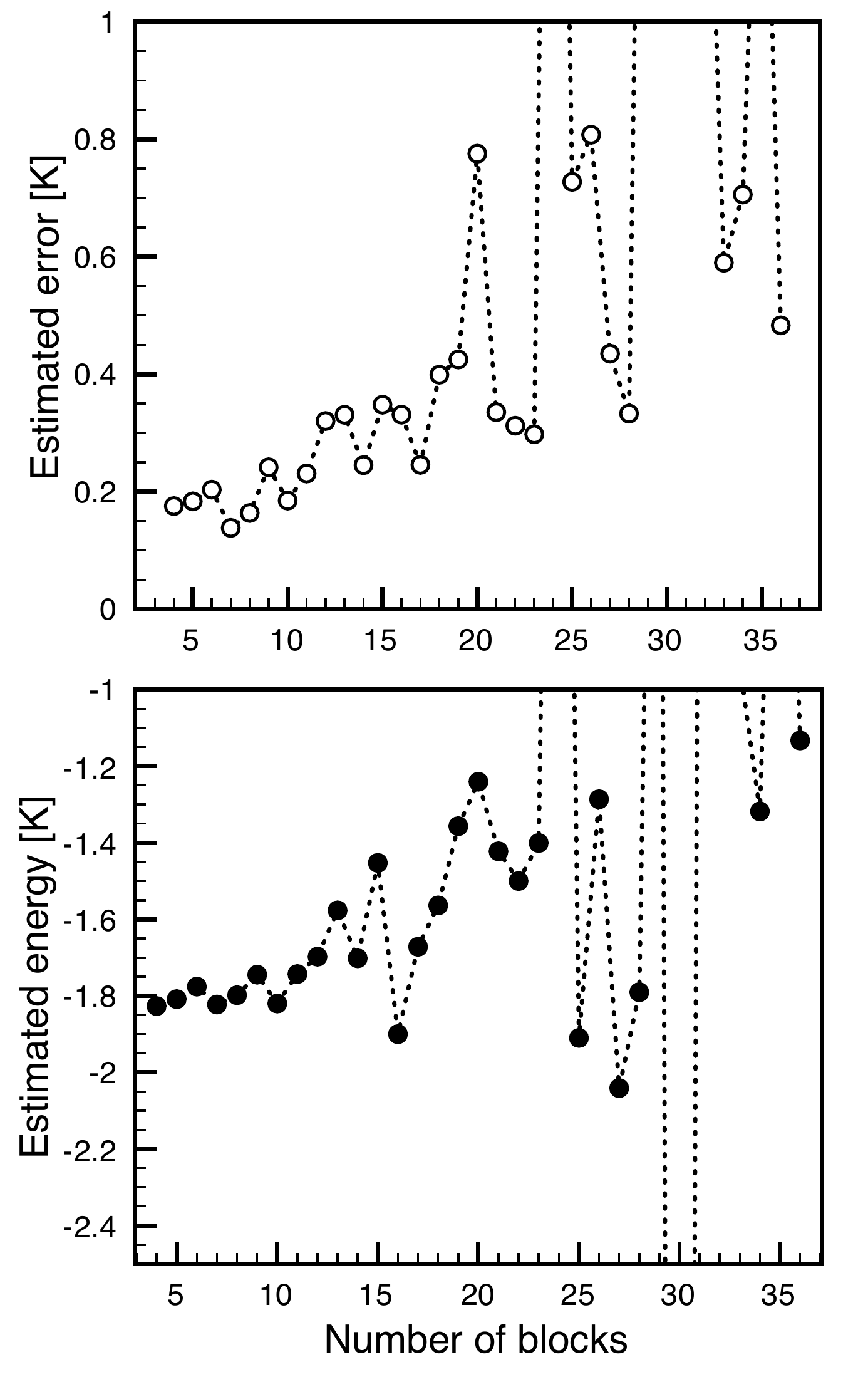}
	\caption{The block average-energy $\langle E \rangle_{\text{block}}$ and the corresponding error bar $\sigma_{\text{block}}$ from a FSWF simulation of $\hetre$ with $N=66$ atoms and $M=128 \cdot 10^8$ sampling points as a function of $n_{\text{block}}$.} 
	\label{fig:sign_problem}
\end{figure}

The block average-energy $\langle E \rangle_{\text{block}}$ and the corresponding standard deviation $\sigma_{\text{block}}$ from a FSWF simulation of $\hetre$ as a function of $n_{\text{block}}$ are shown in Fig.~\ref{fig:sign_problem}. However, the expected plateau onset of $\sigma_{\text{block}}$ and the estimated energy $\langle E \rangle_{\text{block}}$ can only be observed when the lengths of the individual blocks is rather large. As a consequence, the statistical uncertainty of the mean value $\langle E \rangle_{\text{block}}=1.8(2)~\text{K}$ is relatively large, 
which is a clear manifestation of the sign problem of the FSWF. Nevertheless, it has to be said that the present example represents a worst-case scenario for the FWSF and that the sign problem is in this case particularly severe. In fact, if the parameter $C$ of Eq.~\ref{ups} is large, 
$\mathbf{S}_1$ and $\mathbf{S}_2$ are confined around $\mathbf{R}$, which causes that $\text{SD}(\mathbf{S}_1)$ as well as $\text{SD}(\mathbf{S}_2)$ have actually the same sign.

In any case, it is important to emphasize that the sign problem of the FSWF differs from the infamous fermion sign problem of projection methods such as Green's Function \cite{Kalos:1974kx} or Diffusion Monte Carlo \cite{Ceperley:1980vn}. Whereas the latter is conjectured to be nondeterministic polynomial hard \cite{PhysRevLett.94.170201}, which is due to the intrinsic difficulty to sample from a positive pdf generated by a squared antisymmetric function, there is no fundamental reason that prohibits to solve the former and to evaluate a largely fluctuating integral. Nevertheless, the convergence is drastically reduced, so that in many cases, such as the one we have just illustrated, it is virtually impossible to obtain reliable results for any reasonable large number of particles. Apart from that, we would like to point out that an antisymmetric component in the integral always entails convergence problems using MC techniques, so that the present sign problem can be viewed as a particular case of a more general class of integrals.

\section{Antithetic Variates}
\label{sec:antithetic_contributions}
In order to accelerate the convergence, one can examine the behavior of the integrand, and sum over those contributions that lead immediately to a better approximation of the average \cite{AntitheticVariates}. The following example is intended to clarify this concept. 

Suppose that we are interested in numerically evaluating the integral
\begin{equation}
\label{sx01}
I = \int_0^{\infty} dx \, \underbrace{a e^{-ax}}_{f(x)} \underbrace{\sin(\pi x)}_{g(x)} = \frac{\pi a}{\pi^2 + a^2}.
\end{equation}
A MC procedure is to sample $x_k$ from the pdf $f(x_k)$ and form 
the average of $g(x_k)$.  The variance of this estimator reads as 
\begin{eqnarray}
\label{sx02}
var &=&  \int f(x) g^2(x) dx -I^2 \NN
&=& \frac{\pi^2(2\pi^4 + a^4)}{(4\pi^2 + a^2)(\pi^2 + a^2)^2}, 
\end{eqnarray}
whereas the quotient
\begin{eqnarray}
\label{sx03a}
\frac {I^2}{var} &=& \frac{a^2(4\pi^2 + a^2)}{(2\pi^4 + a^4)}
\end{eqnarray}
is a measure of the ``signal-to-noise" ratio and approaches 
\begin{eqnarray}
\label{sx03b}
2\frac{a^2}{\pi^2} + O\left(\frac{a^4}{\pi^4}\right) {\textrm{ as }} a \rightarrow 0.
\end{eqnarray}
That is to say that the procedures becomes very inefficient if $a$ small. 
Since in this case the mean value of $x$, which equals to $1/a$, is large, the 
cancellation of positive and negative lobes of the $\sin$ function becomes 
more pronounced.  This is a simple example of a  ``sign problem''. 

Nevertheless, the problem can be eliminated completely by various
forms of correlated estimates, e.g. by correlating a negative
lobe with the previous (positive but bigger) lobe. The most
effective correlation (and easiest to analyze) is to sample $x$ only
on $[0,1]$, but include all $x+n$, where $n=0\dots \infty$, with the factor
$(-1)^n e^{-na}$. 
Since
\begin{equation}
\label{sx04}
\sum_{n=0}^{\infty} (-1)^n e^{-na} = \frac{e^a}{1+e^a}, 
\end{equation}
we can recast the integral as 
\begin{eqnarray}
\label{sx05}
I &=& \int_0^{1} dx \, \underbrace{\left \{ \frac{a e^{-ax}}{1-e^{-a}} \right \}}_{f_{C}(x)}
 \underbrace{\left \{ \frac{e^a-1}{e^a + 1} \right \} \sin(\pi x)}_{g_{C}(x)} \NN
 &=&  \frac{\pi a}{\pi^2 + a^2}, 
\end{eqnarray}
where again $f_C(x)$ and $g_C(x)$ are the functions to sample and average over, respectively. 
The associated variance reads as 
\begin{eqnarray}
\label{sx07}
var_C &=& \int_0^1 \left \{ \frac{a e^{-ax}}{1-e^{-a}} \right \} 
 \left [  \left \{ \frac{e^a-1}{e^a + 1} \right \} \sin(\pi x) \right ]^2 dx - I^2 \NN
 &=& \frac{2 \pi ^2 \tanh ^2\left(\frac{a}{2}\right)}
 {a^2+4 \pi ^2}-\frac{\pi ^2 a^2}{\left(a^2+\pi ^2\right)^2}
\end{eqnarray}
and the signal-to-noise ratio 
\begin{equation}
\label{sx08}
\lim_{a \rightarrow 0} \frac {I^2}{var_C} = \frac{8}{\pi^2 - 8} +
\frac{2(2\pi^2-21)}{3(\pi^2 - 8)^2}a^2 + O(a^4)
\end{equation}
is now finite, at variance to Eq.~\ref{sx03b}.

In the case of the FSWF, the underlying idea is that the integral
\begin{equation}
	\int d\mathbf{S} \, \Xi(\mathbf{R},\mathbf{S}) \text{SD}(\mathbf{S}) J_s(\mathbf{S})
\end{equation}
has both positive and negative contributions, and that summing pairs of positive and negative values speeds up the convergence. To that extent in the following two promising geometrical transformations that take advantage of the antithetic contributions are proposed: permutations and reflections.

\subsection{Permutations}
\label{sub:permutations}
In all of the presented methods belonging to this category, pair permutations of the shadows are employed to induce the desired antithetic contributions.

\subsubsection{Gaussian Determinant}
\label{ssub:use_of_the_gaussian_determinant}
The first approach is to directly sum over all permuted terms, which eventually translates into a determinant consisting of Gaussians. To illustrate this we sum, on the one hand, over all pair permutations $\mathcal{P}^{(2)}_{ij}$, which leads to 
\begin{eqnarray}
  \psi_{\text{\tiny FSWF}}(\mathbf{R}) &=& J_p(\mathbf{R}) \int d\mathbf{S} \, \Big[\Xi(\mathbf{R},\mathbf{S}) \\ 
  &-& \sum_{i=1}^{N-1} \sum_{j=i+1}^{N} \Xi(\mathbf{R},\mathcal{P}^{(2)}_{ij} \mathbf{S})\Big] \text{SD}(\mathbf{S}) J_s(\mathbf{S}), \nonumber
\end{eqnarray} 
and, on the other hand, over all $3$-term permutations $\mathcal{P}^{(3)}_{ijk}$, i.e.
\begin{eqnarray}
  \psi_{\text{\tiny FSWF}}(\mathbf{R}) &=& J_p(\mathbf{R}) \int d\mathbf{S} \, \Big[\Xi(\mathbf{R}, \mathbf{S}) \\ 
  &+& \sum_{i=1}^{N-2} \sum_{j=i+1}^{N-1} \sum_{k=j+1}^{N} \Xi(\mathbf{R},\mathcal{P}^{(3)}_{ijk} \mathbf{S}) \Big] \text{SD}(\mathbf{S}) J_s(\mathbf{S}). \nonumber
\end{eqnarray}

This is to say that in general an even number of permutations results in a change of sign, while an odd number does not. It is now clear that the sum over all the possible permutations can be elegantly expressed as a matrix determinant $\text{det}\left(e^{-C (\mathbf{r}_{\alpha}-\mathbf{s}_{\beta})^2}\right)$ that we will refer to as Gaussian determinant $\text{GD}(\mathbf{R},\mathbf{S})$. As a consequence, 
\begin{eqnarray}
  \psi_{\text{\tiny FSWF}}(\mathbf{R}) &=& J_p(\mathbf{R}) \int d\mathbf{S} \, \text{GD}(\mathbf{R},\mathbf{S})
  \text{SD}(\mathbf{S}) J_s(\mathbf{S}),
\end{eqnarray}
where $\alpha$ and $\beta$ denotes the matrix rows and columns, respectively.
This representation is particularly convenient, because, similar to the Slater determinant, it permits the summation over all $N!$ terms with a computational cost of $\mathcal{O}(N^3)$, where $N$ is the number of atoms \cite{PhysRev.34.1293}. 

\begin{figure}
\begin{center}
\includegraphics[width=8.5cm]{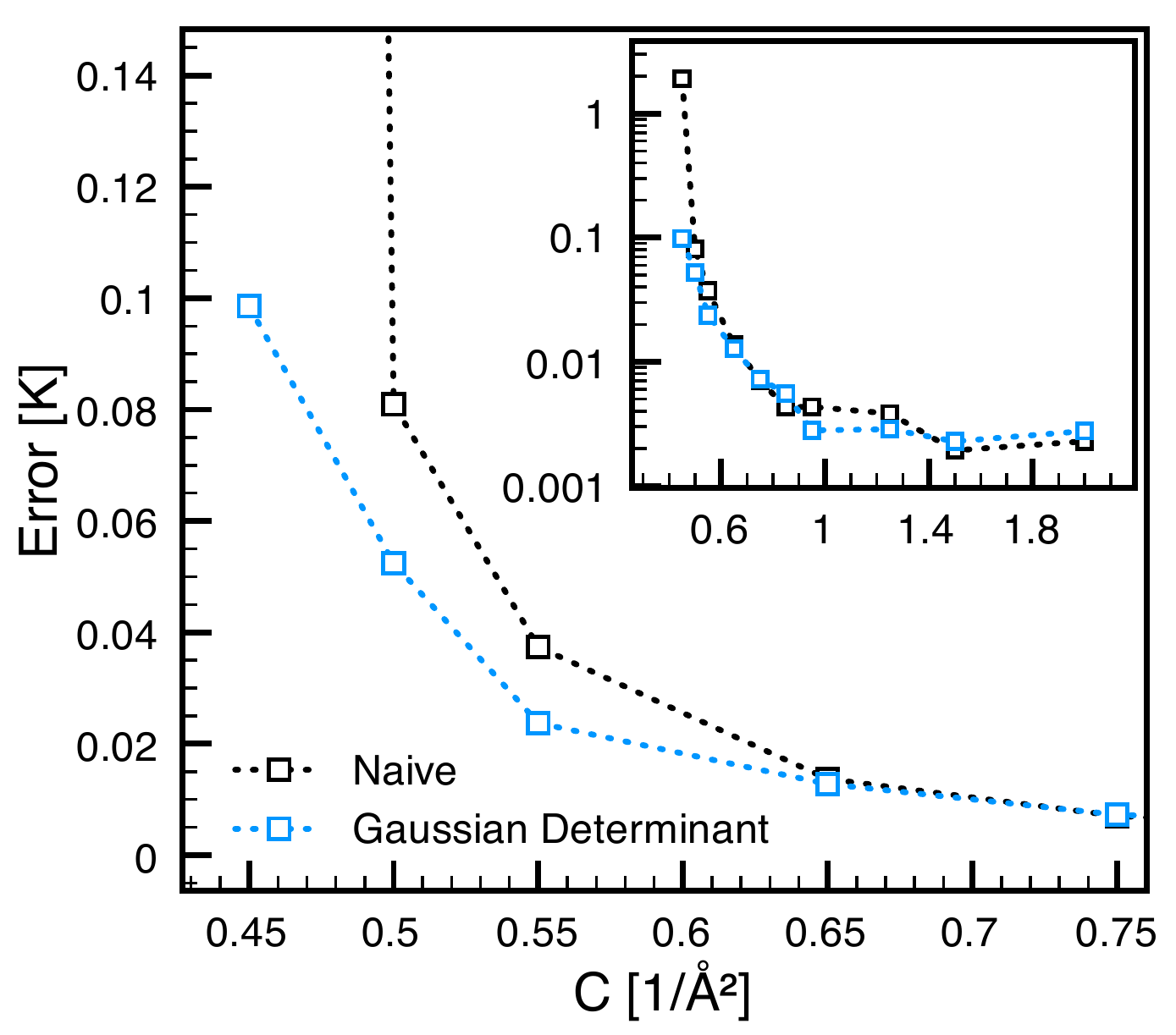}
\end{center}
\caption{\label{imm:gd_interactive_fermion}Estimated error with and without the adoption of the $\text{GD}$ approach, for different values of $C$. Results were obtained for $N=38$ and $M=64 \cdot 10^7$.}
\end{figure}
In other words, in this first scheme, the Gaussian product $\Xi(\mathbf{R},\mathbf{S})$ is replaced by the Gaussian determinant $\text{GD}(\mathbf{R},\mathbf{S})$. The corresponding results are illustrated in Fig.~\ref{imm:gd_interactive_fermion}. 
Due to the fact that the parameter $C$ of Eq.~\ref{ups} is related to the mutual confinement of the particles and shadows, it is large in the crystalline phase, while for a liquid it is relatively small. 
As expected, the $\text{GD}$ method reduces the variance in particular for small values of $C < 0.45~\text{\AA}^{-2}$, whereas in realistic simulations $C$ is typically around $0.5~\text{\AA}^{-2}$ \cite{PhysRevB.53.15129}. 
Overall, the $\text{GD}$ technique throughout reduces the variance although not to the extend to facilitate large-scale calculations without excessive sampling.
However, a potential limitation of the present scheme may arise due to sampling a sum of permuted terms, such that only one of them is sampled efficiently, regardless of all the others.

\subsubsection{Explicit pair permutation term: Duet and Quartet}
\label{ssub:explicit_pair_permutation_term}
Therefore, an alternative approach is to add the contributions that are due to a single pair permutation $\mathcal{P}_{a b}$ to the original integrand, so as to
\begin{eqnarray}
	\psifswf(\mathbf{R}) &=& \int d\mathbf{S} \, \left( \Gamma(\mathbf{R},\mathbf{S}) + \Gamma(\mathbf{R},\mathcal{P}_{a b} \mathbf{S}) \right) \nonumber \\
	&=& J_p(\mathbf{R}) \int d\mathbf{S} \, J_s(\mathbf{S}) \text{SD}(\mathbf{S}) \nonumber \\
	&\times& \left( \Xi(\mathbf{R}, \mathbf{S}) - \Xi(\mathbf{R}, \mathcal{P}_{a b} \mathbf{S}) \right).
\end{eqnarray}
Compared with the $\text{GD}$ method, the latter has the advantage of allowing to sample from a product of permuted terms, instead of a sum:
\begin{eqnarray}
	\rho(\mathbf{R},\mathbf{S}_1,\mathbf{S}_2) &=& \sqrt{\Gamma(\mathbf{R},\mathbf{S}_1) \Gamma(\mathbf{R},\mathcal{P}_{a b} \mathbf{S}_1)} \nonumber \\ 
	&\times& \sqrt{\Gamma(\mathbf{R},\mathbf{S}_2) \Gamma(\mathbf{R},\mathcal{P}_{a b} \mathbf{S}_2)} \label{pair_permutation_sampling_function}
\end{eqnarray}
At variance to the GD, the original as well as the permuted configuration are given equal importance, so that eventually their contributions will be of the same order, which results in a more effective mutual annihilation. The permuted term can either be added to $\Gamma(\mathbf{R}, \mathbf{S}_1)$ alone or to both $\Gamma(\mathbf{R}, \mathbf{S}_1)$ and $\Gamma(\mathbf{R}, \mathbf{S}_2)$, respectively. Due to the fact that this results in two or four terms in the expression for the energy, we will refer to these schemes as the Duet and Quartet techniques, respectively.

To assess the effectiveness this concept, we have performed a calculation for $N=14$ and $M = 16 \cdot 10^6$ using the Duet technique. However, the fluctuations were so high that was difficult to estimate the error using the algorithm described above. We therefore have to conclude that the sampling function of Eq.~\ref{pair_permutation_sampling_function} seem to be not optimal for integrating either $\Gamma(\mathbf{R}, \mathbf{S})$ or $\Gamma(\mathbf{R}, \mathcal{P}_{a b} \mathbf{S})$. 

As a consequence, the efficiency is inferior than the $\text{GD}$ and even worse using the Quartet technique. This is to say that a permutation alone does not provide an effective antithetic contribution, as otherwise the product of $\Gamma$ with its permuted term would exhibit its maximum in the same region where the function itself has its maximum so that the sampling problem would not have emerged in the first place.

\subsubsection{Permutation move}
\label{ssub:permutation_move}
As we have just seen, even if a permutation implies a sign change, it is not necessarily resulting in an optimal antithetic contribution, which is due to the presence of the kernel that breaks the symmetry. 

We can therefore infer that after performing a permutation, a specific translation needs to be added in order to obtain an effective antithetic contribution leading to mutual cancelation. Even though this translation is evidently unknown, it is yet possible to allow a walker to diffuse after a permutation, so that it can spontaneously move to the correct antithetic point. To implement this idea we need to consider permutations as proposed moves for the walkers in the context of the \mrtwo algorithm \cite{MRT}. To that extend, it is of upmost importance to take all possible pair permutations into account, and to select the most favorable one in order to maximize the acceptance rate.

Specifically, given a certain $\mathbf{R}$ and $\mathbf{S}$, we evaluate $\Xi(\mathbf{R},\mathcal{P}_{ij} \mathbf{S})$ for all the possible $i$ and $j$ and propose a permutation $(a,b)$ according to the transition probability
\begin{equation}
	T(\mathbf{S} \rightarrow \mathcal{P}_{ab} \mathbf{S}) = \frac{\Xi(\mathbf{R},\mathcal{P}_{ab})}{\sum_{(i,j)}^{} \Xi(\mathbf{R},\mathcal{P}_{ij})}.
\end{equation}
Thereafter, the acceptance probability has to be modified and reads as
\begin{equation}
	A (\mathbf{S} \rightarrow \mathcal{P}_{ab} \mathbf{S}) = \frac{|\Gamma(\mathbf{R},\mathcal{P}_{ab} \mathbf{S})| T(\mathcal{P}_{ab} \mathbf{S} \rightarrow \mathbf{S})}{|\Gamma(\mathbf{R}, \mathbf{S})| T(\mathbf{S} \rightarrow \mathcal{P}_{ab} \mathbf{S})}.
\end{equation}
The permutation moves are proposed always after the evaluation of the estimator, in order to allow the walkers to diffuse before the next evaluation.

Following this procedure for $N=14$ and $M = 16 \cdot 10^6$, we obtained $E = -1.997(11)~\text{K}$, which has to be compared to $E = -1.986(18)~\text{K}$ using the naive algorithm. These values differs from those of Table~\ref{table:risultati_fswf} for $N=66$ due to the presence of single-particle finite size effects. The acceptance rate for the permutation moves was roughly $2.5\%$. This implies that it is actually possible to employ permutation moves, since the acceptance rate is significant, and that they indeed systematically reduce the variance. But, although our novel permutation moves lower the variance, this is largely due to the smaller correlation between successive steps, and thus cannot be a definitive solution to the sign problem, for which negative correlation factors ought to be introduced.

\subsection{Reflections}
\label{sub:reflections}
\begin{figure}
\begin{center}
\includegraphics[height=3cm]{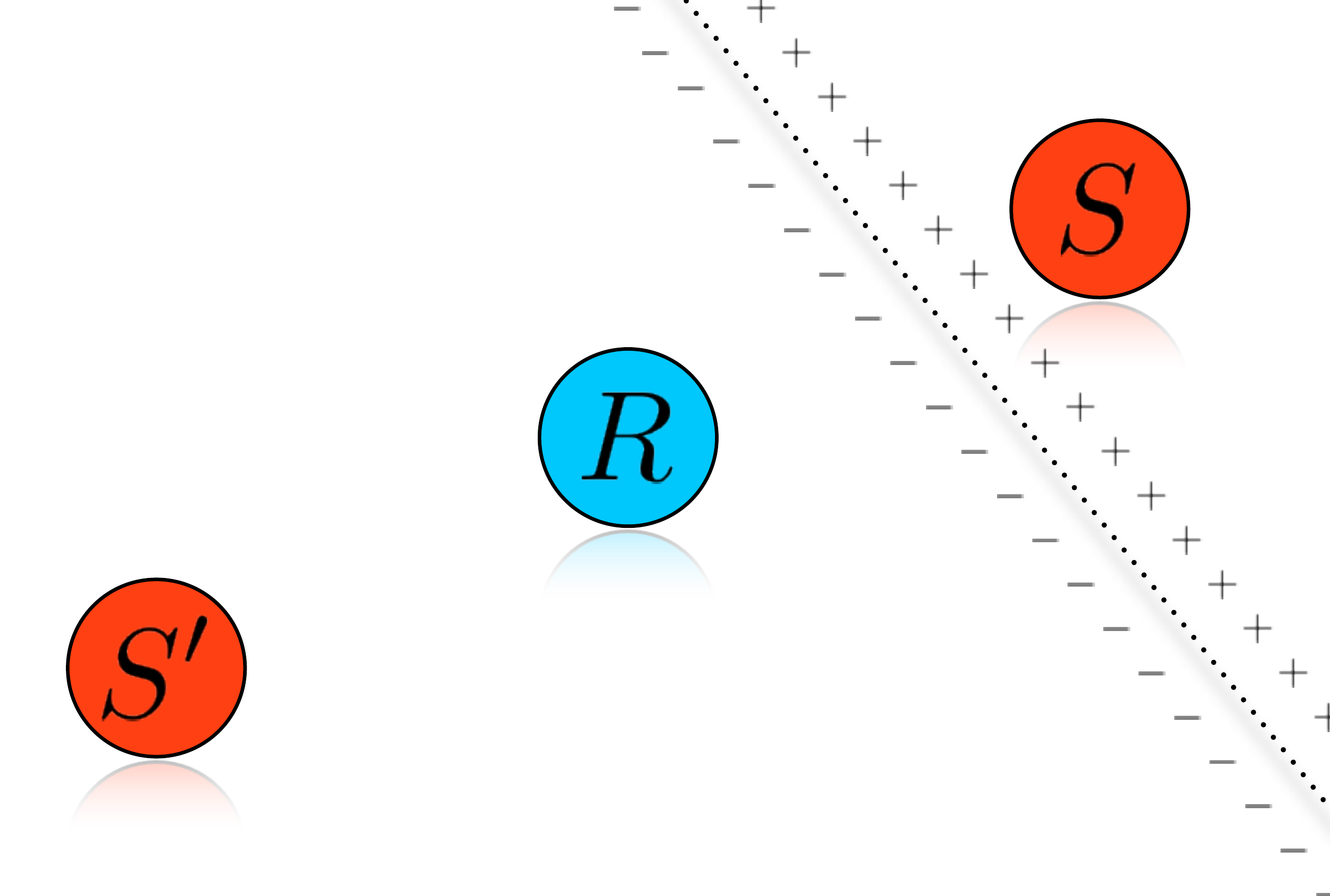}
\end{center}
\caption{\label{imm:reflected_shadow_pdf}Illustration of a reflected shadow $\mathbf{S}^{\prime}$. Since $\mathbf{R}$ is close to the nodal surface defined by the SD, $\mathbf{S}$ and $\mathbf{S}^{\prime}$ are in different nodal pockets.}
\end{figure}

In general, the integral over $\mathbf{S}$ is centered around $\mathbf{R}$ by the Gaussian term. This means that if the parameter $C$ is small, there will be a significant delocalization and $\mathbf{S}$ is more likely to cross the nodal surface defined by the Slater determinant. However, if we additionally also consider the contributions that are arising from the \emph{reflected shadow} $\mathbf{S}^{\prime} = 2\mathbf{R}-\mathbf{S}$, we will possibly obtain an opposite contribution. The concept of this approach is illustrated in Fig.~\ref{imm:reflected_shadow_pdf}.

To take advantage of this idea, we sample from the usual pdf 
\begin{equation}
	\rho(\mathbf{R}, \mathbf{S}_1, \mathbf{S}_2) = | \Gamma(\mathbf{R}, \mathbf{S}_1) \Gamma(\mathbf{R}, \mathbf{S}_2) |, 
\end{equation}
but sum the contributions that are originating from the reflected shadows in the energy estimator
\begin{widetext}
\begin{equation}
	E = \frac{\frac{1}{2} \sum_{i=1}^{M} \left\{ \left[ \frac{ \left( \Gamma(\mathbf{R}, \mathbf{S}_1)+\Gamma(\mathbf{R}, \mathbf{S}^{\prime}_1) \right) \text{H} \left( \Gamma(\mathbf{R}, \mathbf{S}_2)+\Gamma(\mathbf{R}, \mathbf{S}^{\prime}_2) \right)}{|\Gamma(\mathbf{R}, \mathbf{S}_1) \Gamma(\mathbf{R}, \mathbf{S}_2)|} + \frac{ \left( \Gamma(\mathbf{R}, \mathbf{S}_2)+\Gamma(\mathbf{R}, \mathbf{S}^{\prime}_2) \right) \text{H} \left( \Gamma(\mathbf{R}, \mathbf{S}_1)+\Gamma(\mathbf{R}, \mathbf{S}^{\prime}_1) \right)}{|\Gamma(\mathbf{R}, \mathbf{S}_1) \Gamma(\mathbf{R}, \mathbf{S}_2)|} \right] \right\}_i }{\sum_{i=1}^{M} \left\{ \frac{ \left( \Gamma(\mathbf{R}, \mathbf{S}_1)+\Gamma(\mathbf{R}, \mathbf{S}^{\prime}_1) \right) \left( \Gamma(\mathbf{R}, \mathbf{S}_2)+\Gamma(\mathbf{R}, \mathbf{S}^{\prime}_2) \right) }{|\Gamma(\mathbf{R}, \mathbf{S}_1) \Gamma(\mathbf{R}, \mathbf{S}_2)|} \right\}_i }.
\end{equation}
\end{widetext}

But, as it turned out, in a realistic calculation $\Gamma(\mathbf{R}, \mathbf{S}^{\prime})$ is throughout considerably smaller than $\Gamma(\mathbf{R}, \mathbf{S})$, which implies that its contribution is essentially negligible. Specifically, using $14$ particles, the estimated energy and error, with and without the adoption of the $\text{GD}$ method, improved by less a factor of $10^{-4}$. This marginal enhancement immediately suggests that our initial conjecture to generate antithetic contributions by employing reflected shadows needs to be reconsidered. Finally, we remark that the presented reflection method suffers from an infinite variance problem, which can be effectively eliminated by removing the zero values from the sampling function.

\subsection{Constrained Domains}
\label{sub:constrained_domain}
Eventually, it is possible to use symmetry arguments to constrain the domain of the integrals over $\mathbf{R}$, $\mathbf{S}_1$ and $\mathbf{S}_2$, which results in a significant reduction of the integration space.
First of all, due to the antisymmetry requirement of the FSWF, $\mathbf{R}$ is constrained to the positive (or negative) domain of the corresponding $\text{SD}(\mathbf{R})$. No particular form is required for $\text{SD}(\mathbf{R})$: the present method is correct independent of its choice. The second symmetry argument is only valid in conjunction with the $\text{GD}$ method. In fact, if we sum over all the permutations of $\mathbf{S}$, it is possible to integrate $\mathbf{S}_1$ and $\mathbf{S}_2$ only in the positive (or negative) domains, i.e. where $\text{SD}(\mathbf{S}_1)$ and $\text{SD}(\mathbf{S}_2)$ are positive (or negative).
We stress that these restrictions do not imply that $\Gamma$ must always be positive (or negative), as the Gaussian determinant permits a change of sign. 

A simulation using the above described constrained domains method, with $N=38$ and $M = 60 \cdot 10^6$, yielded $E = -2.5(10)~\text{K}$, whereby $5\%$ and $0.5\%$ of the moves for $\mathbf{R}$ and $\mathbf{S}$, respectively, were rejected due to the constraints. Comparing this with $E = -1.9(1)~\text{K}$ using the bare $\text{GD}$ technique without any restrictions, it is clear that no reduction of the sign fluctuations has been achieved. However, the fact that error increases by a factor of $10$ is surprising, but might be explained to be most likely a consequence that when integrating on a restrained domain the efficiency decreases near its borders.

\section{The Grouping Technique and the Marginal Distribution}
\label{sec:the_grouping_technique}
A completely different approach can be devised by analyzing the expression for the energy with a special attention on the integrals over $\mathbf{S}_1$ and $\mathbf{S}_2$:
\begin{equation}
	E = \frac{\int d\mathbf{R} \, \left( \int d\mathbf{S}_1 \, \Gamma(\mathbf{R}, \mathbf{S}_1) \right) \left( \int d\mathbf{S}_2 \, \text{H} \Gamma(\mathbf{R}, \mathbf{S}_2) \right)}{\int d\mathbf{R} \, \underbrace{\left( \int d\mathbf{S}_1 \, \Gamma(\mathbf{R}, \mathbf{S}_1) \right)}_{\Omega_1(\mathbf{R})} \underbrace{\left( \int d\mathbf{S}_2 \, \Gamma(\mathbf{R}, \mathbf{S}_2) \right)}_{\Omega_2(\mathbf{R})} }
\end{equation}
We point out that $\Omega_1(\mathbf{R})=\Omega_2(\mathbf{R})=\psifswf(\mathbf{R})$. From this it follows that knowing $\psifswf(\mathbf{R})$, i.e. knowing the analytical solution of the integral over $\mathbf{S}$, the sign problem ceases to exist, since $\Omega_1(\mathbf{R}) \Omega_2(\mathbf{R}) = \psifswf^2(\mathbf{R}) \ge 0$. The fact that $\psifswf(\mathbf{R})$ is apparently unknown has the following two major consequences. 
First, $\Omega_1(\mathbf{R})$ and $\Omega_2(\mathbf{R})$ needs to be approximated by sampling $\Gamma(\mathbf{R}, \mathbf{S}_1)$ and $\Gamma(\mathbf{R}, \mathbf{S}_2)$, respectively. Due to the fact that the estimates may have have different signs, the local energy is weighted by an extremely noisy function. Second, $\mathbf{R}$ is not efficiently sampled from its marginal distribution $\psifswf^2(\mathbf{R})$. The following example is intended to illustrate this subtle point. Suppose that $\mathbf{R}_0$ is on the nodal surface, i.e. $\psifswf(\mathbf{R}_0)=0$. Even though this configuration should never be sampled since its probability is identical to zero by its very definition, it will, nevertheless, be sampled with a finite probability of $\Gamma(\mathbf{R}_0, \mathbf{S}_1) \Gamma(\mathbf{R}_0, \mathbf{S}_2)$, which is a manifestation that that integrand and the integral may greatly differ from each other. 
Moreover, if $\psifswf \simeq 0$, $\Gamma(\mathbf{R}, \mathbf{S}_1)$ and $\Gamma(\mathbf{R}, \mathbf{S}_2)$ are contributing only with noise.

The first problem has been previously addressed in a study of the vacancy formation energy in solid $\hetre$ using the grouping technique that is based on the aforementioned blocking scheme \cite{dandrea:fswf}. The main idea we have adopted here in our improved grouping technique is to sample many successive shadows in order to obtain a rough estimate of $\Omega_1(\mathbf{R})$ and $\Omega_2(\mathbf{R})$. 
The present algorithm then reads as:
\begin{enumerate}
	\item Start from a configuration $\mathbf{R}_{(0)}$, ${\mathbf{S}_1}_{(0,1)}$, as well as ${\mathbf{S}_2}_{(0,1)}$ and set $i=1$
	\item \label{enum:start_mc_group} Sample $\mathbf{R}_{(i)}$ from the pdf $$\rho(\mathbf{R})=|\Gamma(\mathbf{R}, {\mathbf{S}_1}_{(i-1,1)}) \Gamma(\mathbf{R}, {\mathbf{S}_2}_{(i-1,1)})|$$
	\item Sample $M_s$ points $\left( {\mathbf{S}_1}_{(i,1)}, \dots, {\mathbf{S}_1}_{(i,M_s)} \right)$ from $$\rho(\mathbf{S}_1)=|\Gamma(\mathbf{R}_{(i)}, \mathbf{S}_1) \Gamma(\mathbf{R}_{(i)}, {\mathbf{S}_2}_{(0,1)})|$$ and analog $\left( {\mathbf{S}_2}_{(i,1)}, \dots, {\mathbf{S}_2}_{(i,M_s)} \right)$ from the pdf $$\rho(\mathbf{S}_2)=|\Gamma(\mathbf{R}_{(i)}, {\mathbf{S}_1}_{(i,M_s)}) \Gamma(\mathbf{R}_{(i)}, \mathbf{S}_2)|$$
	\item Evaluate $\Omega_1(\mathbf{R}_{(i)}) = \sum_{j=1}^{M_s} \frac{\Gamma(\mathbf{R}_{(i)}, {\mathbf{S}_1}_{(i,j)})}{|\Gamma(\mathbf{R}_{(i)}, {\mathbf{S}_1}_{(i,j)})|}$ \\ and $\Omega_2(\mathbf{R}_i) = \sum_{j=1}^{M_s} \frac{\Gamma(\mathbf{R}_i, {\mathbf{S}_2}_{(i,j)})}{|\Gamma(\mathbf{R}_i, {\mathbf{S}_2}_{(i,j)})|}$
	\item Evaluate $\hloc_1(\mathbf{R}_{i}) = \sum_{j=1}^{M_s} \frac{\text{H} \Gamma(\mathbf{R}_{(i)}, {\mathbf{S}_1}_{(i,j)})}{|\Gamma(\mathbf{R}_{(i)}, {\mathbf{S}_1}_{(i,j)})|}$ \\ and $\hloc_2(\mathbf{R}_{i}) = \sum_{j=1}^{M_s} \frac{\text{H} \Gamma(\mathbf{R}_{(i)}, {\mathbf{S}_2}_{(i,j)})}{|\Gamma(\mathbf{R}_{(i)}, {\mathbf{S}_2}_{(i,j)})|}$
	\item Set $i=i+1$ \label{enum:end_mc_group}
	\item Repeat the steps \ref{enum:start_mc_group} to \ref{enum:end_mc_group} $M$ times
	\item Compute $E=\frac{ \frac{1}{2} \sum_{i=1}^{M} \left( \Omega_1(\mathbf{R}_i) \hloc_2(\mathbf{R}_i) + \Omega_2(\mathbf{R}_i) \hloc_1(\mathbf{R}_i) \right)  }{ \sum_{i=1}^{M} \Omega_1(\mathbf{R}_i) \Omega_2(\mathbf{R}_i) }$
\end{enumerate}
\begin{table}
\begin{center}
\begin{tabular}{|c|c|}
	\hline
$M_s$             & {Efficiency $\left[\text{sec}^{-1}K^{-2} \right]$} \\ \hline
$1    $                & $0.63$ \\ 
$10   $                & $0.36$ \\ 
$100  $                & $0.36$ \\ 
$1000 $                & $0.23$  \\ \hline
\end{tabular}
\end{center}
\caption{\label{table:summary_results_gt} Efficiency of the improved grouping technique in conjunction with the $\text{GD}$ approach for $N=14$ as a function of the block size $M_s$. The efficiency is defined as $1/(\text{simulation time} \times \text{variance})$.}
\end{table}
We have repeated the calculation using the blocking technique to ensure that successively sampled $\mathbf{R}$ values were actually decorrelated and thus finding the optimal number of diffusive steps $\text{M}_{\text{diff}}$. However, we found that even though the improved grouping technique indeed stabilizes the sign of $\Omega_1(\mathbf{R}) \Omega_2(\mathbf{R})$, the computational time required to do so is not entirely compensated by the reduced variance. This can be seen in Table~\ref{table:summary_results_gt}, where the efficiency that is defined as $1/(\text{simulation time} \times \text{variance})$ is reported for different values of $M_s$. Nevertheless, although this scheme alone does not improve the efficiency for the liquid phase, it potentially does for the solid state, where the sign problem is less severe.

In order to make further progress, we will focus on the second problem, whose solution may also solve the first one \textit{en passant}. In this respect we are going to propose two different methods: In the first approach the marginal distribution is approximated analytically, whereas in the second scheme it is estimated numerically instead.

\subsection{J-SD approximation}
\label{ssub:j_sd_approximation}
In our first approach, the J-SD trial wave function is employed as an approximation for $\psifswf(\mathbf{R})$ to sample $\mathbf{R}$, which is why we call this technique J-SD approximation. 
However, in this way $\mathbf{R}$ would be sampled independent from its shadows that would require to relax them whenever $\mathbf{R}$ changes. To avoid this, we have decided to use the same sampling function for $\mathbf{R}$ and $\mathbf{S}$. 
At variance to the just described algorithm based on the blocking technique, $\Gamma(\mathbf{R}, \mathbf{S})$ is replaced by $\Gamma^{\prime}(\mathbf{R}, \mathbf{S})$ together with appropriate weights in the energy estimator. Specifically, the following forms are proposed here, which all incorporates $\text{SD}(\mathbf{R})$ into the sampling function for $\mathbf{R}$:
\begin{itemize}
	\item $\Gamma_1^{\prime}(\mathbf{R}, \mathbf{S}) = \Gamma(\mathbf{R}, \mathbf{S}) \left( \SD(\mathbf{R})^2 + \Lambda^2(\mathbf{R}) \right)^{1/4}$
	\item $\Gamma_2^{\prime}(\mathbf{R}, \mathbf{S}) = \Gamma(\mathbf{R}, \mathbf{S}) \left( \SD(\mathbf{R})^2 + \Lambda^2(\mathbf{R}) \right)$ 
	\item $\Gamma_3^{\prime}(\mathbf{R}, \mathbf{S}) = \Gamma(\mathbf{R}, \mathbf{S}) \left( \SD(\mathbf{R})^2 + \Lambda^2(\mathbf{R}) \right)^{1/2}$
	\item $\Gamma_4^{\prime}(\mathbf{R}, \mathbf{S}) = \Gamma(\mathbf{R}, \mathbf{S}) \left( \SD(\mathbf{R})^4 + \Lambda^4(\mathbf{R}) \right)^{1/4}$
\end{itemize}
To prevent the infinite variance problem, we have introduced an auxiliary factor $\Lambda(\mathbf{R})$, whose optimal value is expected to be of the same order as $\SD(\mathbf{R})$.

\begin{figure}
\begin{center}
\includegraphics[width=8.5cm]{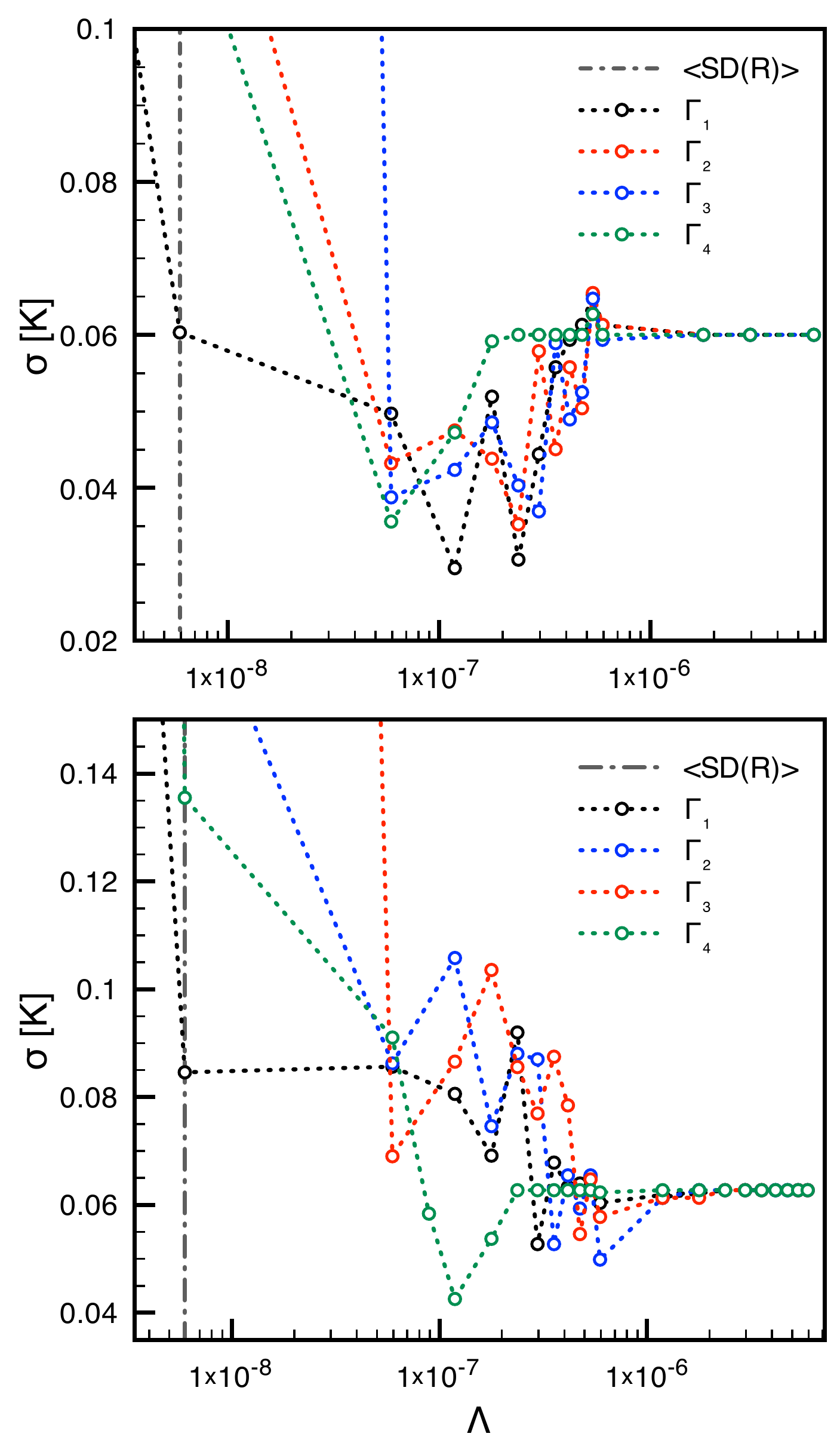}
\end{center}
\caption{\label{imm:sd_r_pdf}Comparison of the error for $N=38$ with respect to $\Lambda(\mathbf{R})$ as obtained using the $\text{GD}$ approach together with the modified grouping technique in conjunction with the J-SD approximation, i.e. $\Gamma^{\prime}(\mathbf{R}, \mathbf{S})$ instead of $\Gamma(\mathbf{R}, \mathbf{S})$. In the upper panel, the blocking technique with $M_s=100$ was used and averaged over $M=96 \cdot 10^4$ points. 
Instead, in the lower panel $M=48 \cdot 10^6$, though without utilizing the blocking technique. The expectation value $\langle \SD(\mathbf{R}) \rangle$ was evaluated by sampling from the pdf $\Gamma(\mathbf{R}, \mathbf{S}_1) \Gamma(\mathbf{R}, \mathbf{S}_2)$.}
\end{figure}
From Fig.~\ref{imm:sd_r_pdf} we can conclude that by incorporating $\text{SD}(\mathbf{R})$ into the sampling function it is possible to reduce the variance by up to a factor of $3/2$. Among the various sampling functions we proposed, $\Gamma_4^{\prime}$ appears to be the most effective, which demonstrates that without the extra term $\mathbf{R}$ is not efficiently sampled from its marginal distribution.

\subsection{S-averaged marginal distribution}
\label{ssub:s_averaged_marginal_distribution}
An alternative possibility that we have investigated here is to employ a numeric estimate of $\Omega_1$ and $\Omega_2$ as the sampling function. 
To that extend we assume $2 N_S$ shadows $\mathbf{S}_{1,1}, \mathbf{S}_{1,2}, \dots, \mathbf{S}_{1,N_S}, \mathbf{S}_{2,1}, \mathbf{S}_{2,2}, \dots, \mathbf{S}_{2,N_S}$, and replace
\begin{subequations}
\begin{equation}
	\int d\mathbf{S}_1 \, \Gamma(\mathbf{R}, \mathbf{S}_1) \quad \text{with} \quad \sum_{i=1}^{N_S} \int d\mathbf{S}_{1,i} \, \Gamma(\mathbf{R}, \mathbf{S}_{1,i}),
\end{equation}
and
\begin{equation}
	\int d\mathbf{S}_2 \, \Gamma(\mathbf{R}, \mathbf{S}_2) \quad \text{with} \quad \sum_{i=1}^{N_S} \int d\mathbf{S}_{2,i} \, \Gamma(\mathbf{R}, \mathbf{S}_{2,i}).
\end{equation}
\end{subequations}
From this it follows that our sampling function will take the form 
\begin{eqnarray}
	& & \rho(\mathbf{R}, \mathbf{S}_{1,1}, \dots, \mathbf{S}_{1, N_S}, \mathbf{S}_{2,1}, \dots, \mathbf{S}_{2, N_S}) = \nonumber \\
	& & \left| \left( \sum_{i=1}^{N_S} \Gamma(\mathbf{R}, \mathbf{S}_{1,i}) \right) \left( \sum_{i=1}^{N_S} \Gamma(\mathbf{R}, \mathbf{S}_{2,i}) \right) \right|.
\end{eqnarray}
In this way, $\mathbf{R}$ is sampled from a more accurate approximation of $\Omega_1(\mathbf{R}) \Omega_2(\mathbf{R})$ than in the J-SD approximation.

\begin{table}
\begin{equation*}
	\begin{array}{|c|c|}
		\hline
		{N_S}	&	\text{Error}	\\
		\hline
		\hline
		1	&	0.024	  \\
		\hline                         
		2	&	0.025	  \\
		\hline                         
		4	&	0.034	  \\
		\hline                         
		10	&	0.029	  \\
		\hline
	\end{array}
\end{equation*}
\caption{\label{table:marginal_distribution} The error for $N=14$ and $M=8 \cdot 10^6$ as obtained by the $\mathbf{S}$-averaged marginal distribution approach in conjunction with the $\text{GD}$ method with respect to the number of shadows $2N_S$.}
\end{table}
\begin{figure}
\begin{center}
\includegraphics[width=8.5cm]{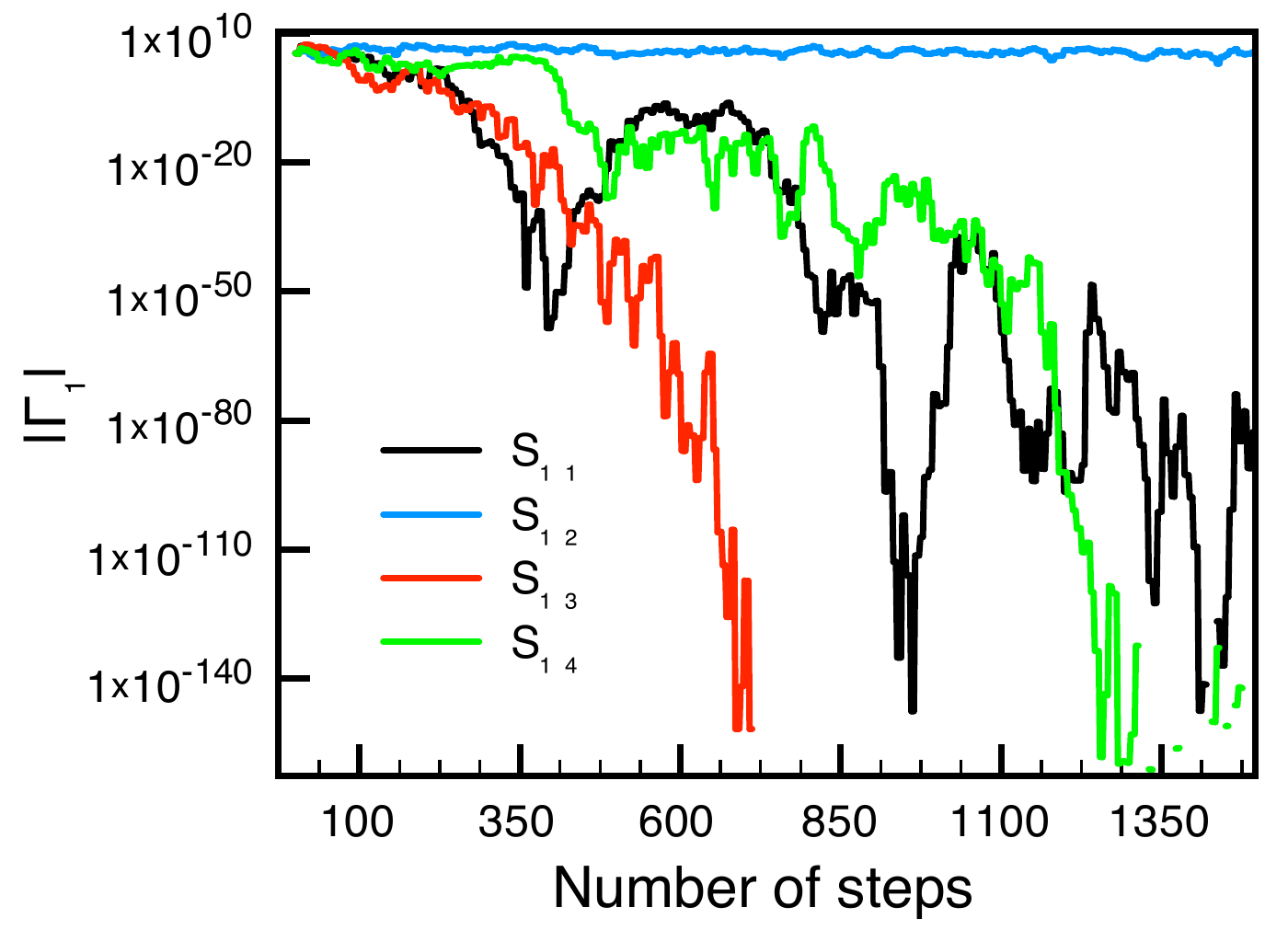}
\end{center}
\caption{\label{imm:falling_shadows_pdf}Trend of the various $\Gamma$ associated to $\mathbf{S}_{1,1}, \dots, \mathbf{S}_{1,4}$, during the progression of a simulation. The red line is interrupted when, because of the double precision float limitations, the value of $\Gamma$ is so small that it is numerically equal to zero and it is no longer visualizable on a logarithmic scale.}
\end{figure}
The results, which are shown in Table~\ref{table:marginal_distribution}, implies that the introduction of additional shadows does not have any statistical significant influence on the variance. From Fig~\ref{imm:falling_shadows_pdf} it is clear why assuming multiple shadows is not effective: During the sampling only one of the $\mathbf{S}_{1,i}$ and one of the $\mathbf{S}_{2,i}$ become significant, whereas all the others tend to zero. As a result, the algorithm returns to its original form, which is hence an inherent consequence of sampling from a sum of $\Gamma(\mathbf{R}, \mathbf{S}_{1,i})$ and $\Gamma(\mathbf{R}, \mathbf{S}_{2,i})$, respectively.

\section{Discussion}
\label{sec:discussion}

The reduced variance of the presented methods, specifically the GD and the J-SD approaches, has to be put in relation with the required computational effort. Thus, for 
the purpose to assess the various techniques presented here, we have summarized their corresponding efficiencies 
in Table~\ref{table:summary_results1}. All the presented results were obtained by means of the \mrtwo algorithm \cite{MRT}, where single-particle or single-shadow moves were proposed at random without any additional drift term. In this process, the step lengths were constantly adjusted to yield an acceptance ratio of $\sim 50 \%$. Due too the fact that the Duet, Quartet, and $R$-$S$ domain constraint methods are apparently inefficient, no error bars for the efficiency were calculated. On the contrary, the J-SD approximation somewhat reduces the variance, though this largely eroded by the additional computational cost. As a consequence, the J-SD method only marginally more efficient. The $\text{GD}$ technique, however, does indeed exhibit a sizable variance reduction. In spite of its increased computational cost to evaluate the Gaussian determinants, it is yet very competitive with the original approach, though generally not significantly more efficient either. Nevertheless, for $C<0.5~\text{\AA}^{-2}$ the $\text{GD}$ method is clearly superior. 
\begin{table}
	\vspace{1cm}
\begin{center}
\begin{tabular}{|c|c|c|}
	\hline
\shortstack{\bf{Used} \\\bf{technique} }              & \shortstack{\bf{Energy} \\ $[\text{K}]$ } & \shortstack{\bf{Efficiency} \\ $[\text{sec}^{-1} \text{K}^{-2}]$}\\ \hline
	\hline
Naive                  & $-1.949 \pm 0.016$ & $30 \pm 3$ \\ \hline
Gaussian determinant   & $-1.943 \pm 0.014$ & $22 \pm 3$ \\ \hline
Duet                   & $2.7 \pm 4.3$ & $\approx 0$ \\ \hline
Quartet                & $-2.2 \pm 1.6$ & $\approx 0$  \\  \hline
Permutation moves      & $-2.012 \pm 0.016$ & $25 \pm 3$ \\ \hline
Reflections$^{\dagger}$           & $-1.945 \pm 0.016$ & $15 \pm 2$ \\ \hline
$R$-$S$ domain constraint$^{\dagger}$ & $-1.974 \pm 0.028$ & $\approx 5$ \\ \hline
$R$ domain constraint$^{\dagger}$   & $-1.975 \pm 0.017$ & $24 \pm 3$ \\ \hline
J-SD approximation      & $-1.953 \pm 0.015$ & $31 \pm 3$  \\ \hline
J-SD approximation$^{\dagger}$     & $-1.949 \pm 0.017$ & $35 \pm 7$  \\ \hline
\end{tabular}
\end{center}
\caption{\label{table:summary_results1}Average energy, associated error and efficiency of all the presented methods. In each case, we have performed 8 independent simulations with $N=38$ and $M =12 \cdot 10^8$. 
For the J-SD approximation method we set $M_s=1$. We did not estimate the efficiency error for certain methods, because they clearly proved unable to provide an improvement. The symbol $^{\dagger}$ denotes that the technique was used in conjunction with $\text{GD}$ method. The efficiency of the Duet and Quartet approaches was extremely low (minor than $10^{-3}$) and therefore approximated to be zero. The $R$-$S$ domain constraint scheme refers to the algorithm described above, whereas for the $R$ domain constraint technique, we limited the constraint only to $R$. For the sake of readability, we multiplied the efficiency by an arbitrary value of 100.}
\end{table}
\begin{figure}
	\centering
		\includegraphics[width=8.5cm]{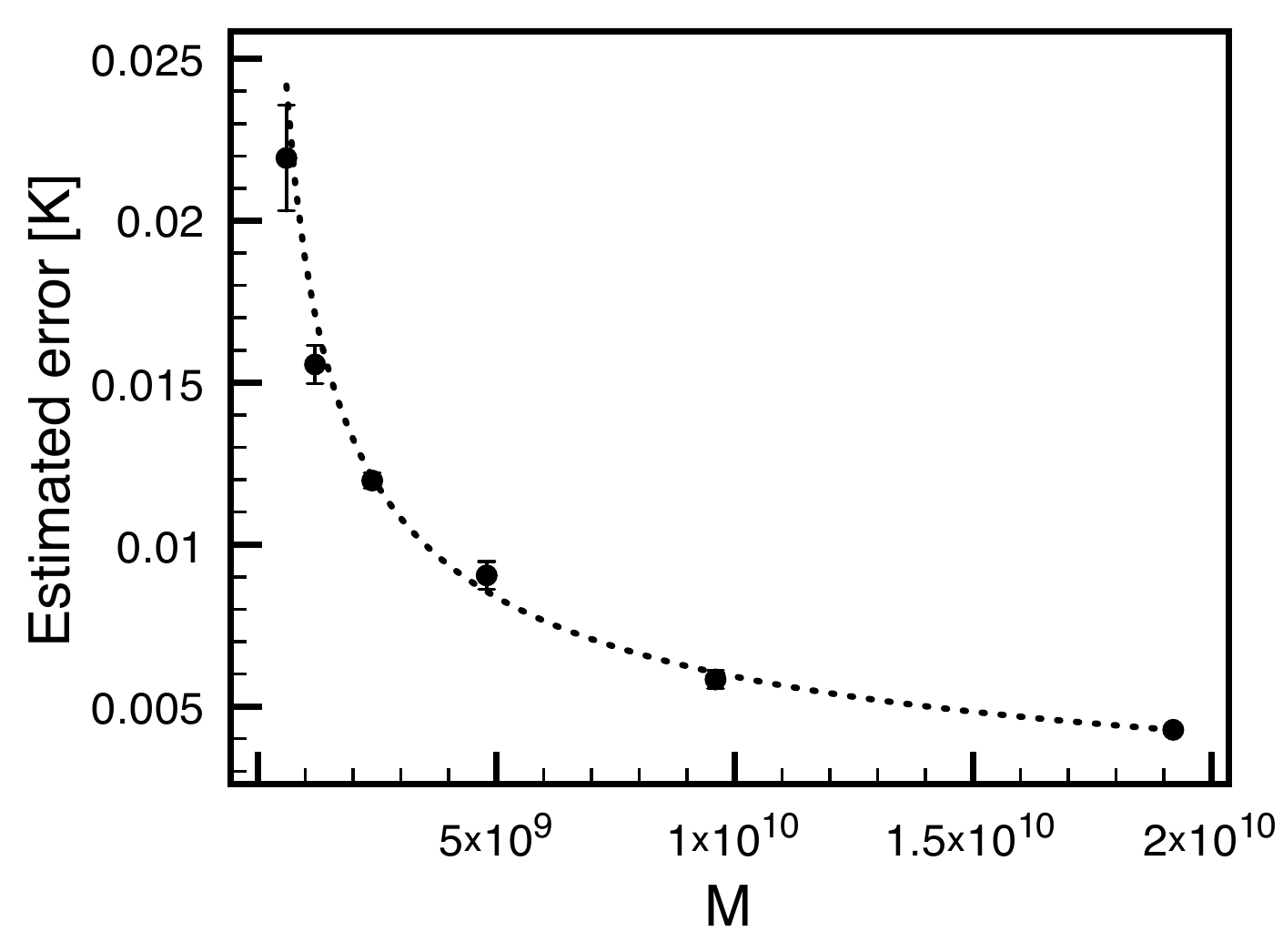}
	\caption{Fitting procedure to estimate the efficiency more accurately. Here we present the data obtained with the naive method, simulating 38 $\hetre$ atoms. Each estimated error was calculated averaging over the eight independent simulations. The fitted function $f(x)=A/\sqrt{M}$ is represented with a dotted line. The chi-squared test was successfully fulfilled.}
	\label{fig:fit}
\end{figure}

Eventually, the combination of the $\text{GD}$ and J-SD approximation methods turned out to be the best among the various technique we have devised here. For this reason we review the accuracy of our results, in order to exclude the possibility that our outcomes were affected by an ergodicity problem. To that extend we have performed several additional calculations for different values of $M$. Then we have fitted the obtained errors to the function $f(x)=A/\sqrt{M}$, which is the expected asymptotic behavior, as shown in Fig.~\ref{fig:fit}. The fact that the chi-squared test was passed successfully indicates that $M$ was large enough to ensure ergodicity. 
In addition, the eventual parameter $A$ can be used as an estimate for the efficiency instead of the variance. The final results are reported in Table~\ref{table:summary_results2}. 

\begin{widetext}
\begin{center}
\begin{table}
	\vspace{1cm}
\begin{tabular}{|c|c|c|c|}
	\hline
\bf{Technique}         & \bf{N=16 - Efficiency} & \bf{N=38 - Efficiency} & \bf{N=54 - Efficiency} \\ \hline
	\hline
Naive                  & $1375 \pm 50$ & $3.03 \pm 0.06$ & $0.0210 \pm 0.0005$ \\ \hline
Gaussian Determinant   & $1220 \pm 30$ & $2.54 \pm 0.14$ & $0.0216 \pm 0.0032$ \\ \hline
J-SD approximation*    & -             & $3.23 \pm 0.22$ &     -               \\ \hline
\end{tabular}
\caption{\label{table:summary_results2} Accurate efficiency estimates of the GD and J-SD techniques, for different number of atoms $N$. We decided to pass over evaluating the efficiency of the "J-SD approximation*" method for $N>38$, since there are no reason to expect significantly different results.}
\end{table}
\end{center}
\end{widetext}

\section{Conclusions}
\label{sec:conclusions}

To summarize, beside revisiting the FSWF and demonstrating the origin and implications of the corresponding sign problem,  
we have proposed two families of novel methods to solve it: Antithetic variates and an improved marginal distribution to sample from. Several specific implementations of these ideas were presented. Even though the GD and J-SD methods are indeed rather effective in reducing the variance, the gain in efficiency is limited due to increased computational cost associated with them. 

We thus conclude that although the presented techniques alleviate the sign problem and allow for very accurate calculations of fermionic systems up to 66 particles, at least when using state of the art supercomputers, a general solution of the sign problem is still outstanding.

\begin{acknowledgments}
  Financial support from the IDEE project of the Carl Zeiss Foundation is kindly acknowledged. T.D.K would like to thank the Graduate School of Excellence MAINZ and the Max-Planck Graduate Center for financial support and the Gauss Center form Supercomputing (GCS) for providing computing time through the John von Neumann Institute for Computing (NIC) on the GCS share of the supercomputer JUQUEEN at the J\"ulich Supercomputing Center (JCS).
\end{acknowledgments}

\bibliographystyle{apsrev}
\bibliography{referenze}

\end{document}